\documentclass{article}
\usepackage{epsfig}
\newcommand{\bfr}{\begin{flushright}}
\newcommand{\efr}{\end{flushright}}
 
\begin{document}
\title{Decaying Domain Walls in an Extended Gravity Model and Cosmology
}
\author{ 
Kiyoshi Shiraishi\\
Institute of Physics, Faculty of Science, Ochanomizu University\\
1-1 Otsuka 2, Bunkyo-ku Tokyo 112, Japan
}
\date{Revista Mexicana de F\'{\i}sica {\bf 38} (1992) 269--278
}
\maketitle
\begin{abstract}
We investigate cosmological consequences of an extended gravity
model which belongs to the same class studied by Accetta and Steinhardt
in an extended inflationary scenario. But we do not worry about
inflation in our model; instead, we focus on a topological object formed
during cosmological phase transitions. Although domain walls appear
during first-order phase transitions such as QCD transition, they decay
at the end of the phase transition. Therefore the ``domain wall problem''
does not exist in the suitable range of pameters and, on the contrary,
the ``fragments'' of walls may become seeds of dark matter. A possible
connection to ``oscillating universe'' model offered by Morikawa et al.
is also discussed.
\end{abstract}

\bigskip

In recent years, many ideas on theory of gravitation have been
produced by lots of authors. From the viewpoint of experimental
physics, the existence of the ``fifth force'' has been examined in these
days \cite{1}. The tests for gravity theory have been managed in
various styles, by means of newly-developed instruments and methods
\cite{2}. From the theoretical viewpoint, many authors have studied
possible modifications and corrections to Einstein's gravity theory,
in consideration of an attempt to unify gravitation and quantum
mechanics. According to superstring theory \cite{3} which has attracted
particle-physicists' attention for several years, Einstein gravity takes
some corrections at very high energy \cite{4}.

 Of course, we cannot discuss gravitation theory apart from evolution
of the universe \cite{5}. Various modifications of gravity theory have
been tried in order to solve cosmological problems. Observational and
theoretical constraints on the amount of time-variation of Newton
constant have been investigated and the upper limit is given \cite{6}.
For planning precise tests of gravity, it is meaningful to study possible
consequences of models which admit spatial and temporal variation of the
effective Newton ``constant''. We can express results of experimental
check for models of gravitation by putting observational constraints on a
few parameters.

A recent idea of ``extended inflation'' scenario \cite{7} includes
an interesting modification of Einstein gravity. In the ``extended
inflation'' scenario, difficulties in ``old'' inflation \cite{8} may be
avoided by introducing a variable Newton ``constant'' into the theory. In
some models of this type, theories of scalar fields coupled non-minimally
to gravity are considered, and they are very similar to Brans-Dicke's
graviy theory \cite{9}. Many variants are also considered in the
literature \cite{10,11,12}.

As a variant of extended inflation model, ``hyperextended inflation''
\cite{10} model is known. In this model, self-interaction of the scalar
field which is coupled non-minimally to gravity, is postulated and a
crucial difference from Brans-Dicke theory is that the scalar field has
an equilibrium point of motion. Owing to this property, one can take a
wide range of values or parameters keeping away from conflicts with
observations.

In the present article, we consider a model of this type which permits
variations of an effective Newton constant in space and time. This model
is a general, and probably simplest model, which can describe the
variation of the strength of gravity. We do not consider inflation
in this paper, however. The reason is that there are too many problems
left unsolved in inflationary models to examine here; they are for
example, the problem of the origin of the very large structure of
galaxies recently discovered \cite{13}, etc.

Throughout this paper, we choose so-called natural units, i.e., we set
$h/(2\pi)=c=k_B=1$.

Our starting point is to consider the followmg Lagrangian
\begin{equation}
{\cal L}=-\frac{1}{2\kappa^2}f(\phi)
R+\frac{1}{2}(\nabla_\mu\phi)^2+\lambda\,.
\label{(1)}
\end{equation}
In this equation, $\phi$ is a neutral scalar field coupled non-minimally
to gravity and $f$ is a function of $\phi$ and gives the manner of
non-linear coupling to gravity; $\kappa^2$ is a constant which has the
same dimensions and order of magnitude as the Newton constant, i.e.,
$\kappa\approx 2.4\times 10^{18} \mbox{GeV}^{-1}$; $\lambda$ stands for
an effective cosmological term which comes from the contribution of the
other matter fields. This term arises, for example, durng a phase
transition of a matter field.

A Weyl transformation of the metric $(g_{\mu\nu}\rightarrow
f^{-1}g_{\mu\nu})$ converts the Lagrangian to Lagrangian including
conventional Einstein-Hilbert term \cite{14}
\begin{equation}
\tilde{{\cal L}}=-\frac{1}{2\kappa^2}f(\phi)
R+\frac{1}{2}\left\{\frac{1}{f}(\nabla_\mu\phi)^2+\frac{3}{2\kappa^2}\left(
\frac{\nabla_\mu f}{f}\right)^2\right\}+\frac{1}{f^2}\lambda\,.
\label{(2)}
\end{equation}
In the following, we investigate the property of the model based on this
Lagrangian (\ref{(2)}); whenever we can describe a phenomenon both in
terms of the Lagrangian (\ref{(1)}) and that of (\ref{(2)}).

If the effective cosmological term $\lambda$ is present, nonlinear self
coupling of the scalar field $\phi$ appears in the Lagrangian
(\ref{(2)}). On account of the self-interaction, topological object in
which energy of the scalar field concentrate can form. Such objects take
the form of ``domain walls'' \cite{15}. We study the properties of the
object and the role in cosmology.

\begin{figure}[ht]
\begin{center}
\includegraphics[width=6cm]{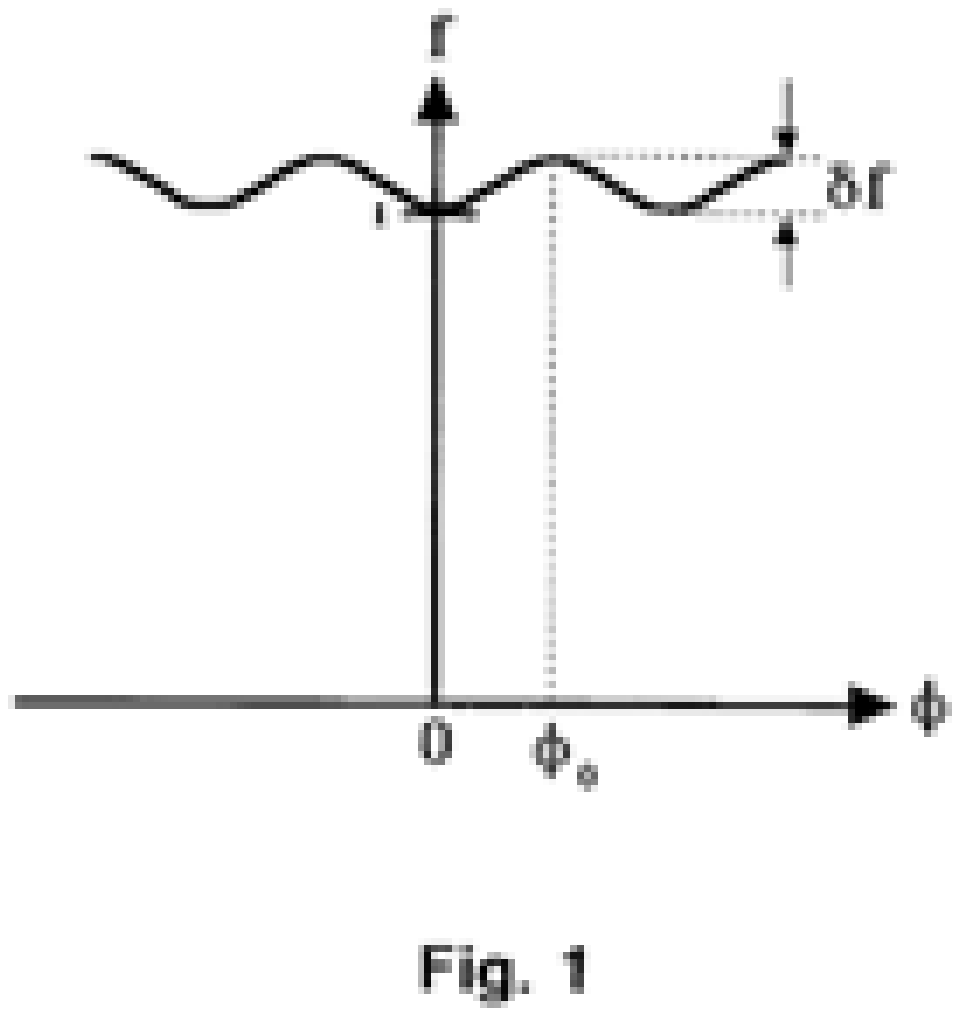}
\caption{A schematic view of the function $f$.}
\label{f1}\end{center}
\end{figure}

First we have to specify the functional form of $f$. We do not expect
drastic changes in the effective strength of gravity in the course of
evolution of the universe.  Accordingly, it is suitable to choose a
periodic continuous function, as a trigonometric function (Fig.~1). The
difference between the maximum and minimum of $f$ is denoted by $\delta
f$, for later use. First of all, we study domain walls; thus $f$ has only
to take the following ``piecewise-linear'' shape (Fig.~2):
\begin{eqnarray}
f(\phi)&=&1+a\left(\phi-\left[\frac{\phi}{\phi_0}\right]\phi_0\right)
\,,\qquad\quad \mbox{if } \left[\frac{\phi}{\phi_0}\right]
\mbox{ is an even integer}\,,\nonumber
\\
f(\phi)&=&1-a\left(\phi-\left[\frac{\phi}{\phi_0}\right]\phi_0-\phi_0
\right)
\,,\quad \mbox{if } \left[\frac{\phi}{\phi_0}\right]
\mbox{ is an odd integer}\,,
\end{eqnarray}
where $[x]$ denotes an integer part of $x$. This $f$ involves two
parameters, $a$ and $\phi_0$. In this approximation, $\delta f$ equals to
$a\phi_0$. In general, $f$ is characterized by $(\delta f)$ and
$(\kappa^2\phi_0^2)$.

\begin{figure}[ht]
\begin{center}
\includegraphics[width=6cm]{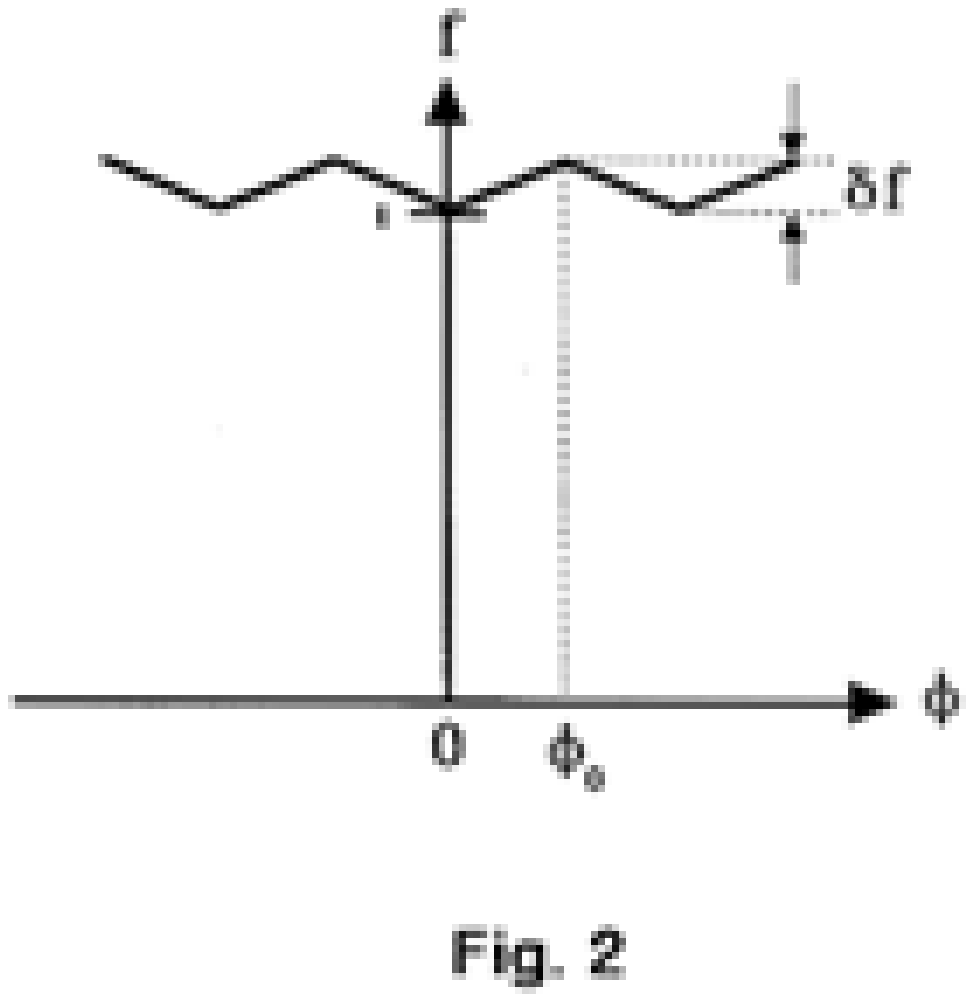}
\caption{The ``piecewise-linear'' approximation of $f$.}
\label{f2}\end{center}
\end{figure}

Suppose that some matter field undergoes first-order phase transition.
During the transition, an effective cosmologlcal term $\lambda$ arises.
Then the neutral scalar field $\phi$ can form a configuration of ``kink''
in one direction, say, in the $z$-axis. In three dimensions, this kink
seems as an infinitely-spread wall of high energy density, parallel to
$x$-$y$ plane \cite{15}. The (domain) wall of this type appears in
general theories with spontaneously broken discrete symetry \cite{15}.
In our model we assume symmetry with respect to $\phi\leftrightarrow
-\phi$.

We want to know qualitative features of the object. We also wish to
avoid a complicated situation where background curvature cannot be
ignored.

To this end, let us suppose that a dimensionless combination
$(\delta f)^2/(\kappa^2\phi_0^2)$ is small in comparison with unity.
Then the following static configuration can be obtained as an approximate
solution (Fig.~3)
\begin{eqnarray}
\phi(z)=h(z)&\equiv&\lambda a z(2x_0-z)\,,\quad (0\le z\le z_0)\,,
\nonumber
\\ &\equiv&\phi_0\,,\qquad\qquad\qquad(z\ge z_0)\,,
\label{(4)}
\end{eqnarray}
where $z_0\equiv\sqrt{\phi_0/\lambda a}=\sqrt{\phi_0^2/\lambda (\delta
f)}$, and it should be read as
$h(-z)=-h(z)$.

\begin{figure}[ht]
\begin{center}
\includegraphics[width=6cm]{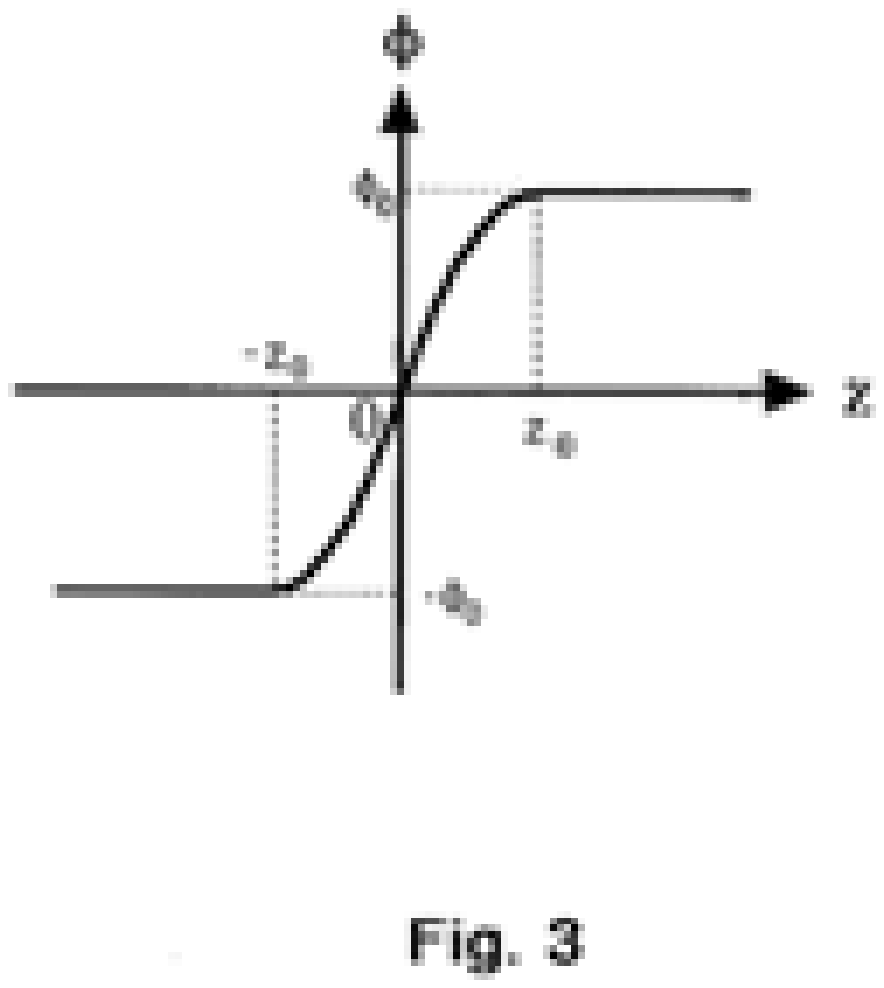}
\caption{The ``kink'' as a static configuration of $\phi$ in the
presence of $\Lambda$.}
\label{f3}\end{center}
\end{figure}

This kink solution is lying in the $z$-direction; thus this
configuration is understood as infinitely stretching domain wall in the
$x$-$y$ plane.

An effective cosmological term entirely vanishes when the phase
transition of the matter field is over. Then, the scalar field loses its
self-interaction term and only has the non-minimal kinetic term. The
domain wall generated during the phase transition is no longer stable
when the phase transition finishes. Now then, how does the wall decay?
We assume plane-symmetric walls here again.

There being no potential term, we have only to solve substantially a
wave equation with respect to an initial condition. We take the static
solution (\ref{(4)}) as an initial condition at the time $t=0$.
Furthermore we assume that the time-derivative of $\phi$ is zero
everywhere, at $t=0$. If the assumption is taken,
the effective Lagrangian (\ref{(2)}) for $\phi$ reduces to a dominant
term (now $\lambda=0$)
\[
\tilde{{\cal L}}\approx\frac{1}{2}\frac{1}{f}(\nabla\mu\phi)^2\,.
\]
Therefore the wave equation we must solve
is approximated as
\begin{equation}
\nabla^\mu\left(\frac{1}{f}\nabla_\mu\phi\right)=0\,. 
\label{(5)}
\end{equation}
Using the kink (\ref{(4)}) as an initial configuration, we can solve the
equation as
\begin{equation} 
\phi(z,t)=\frac{1}{a}\left[\frac{1}{4}\{g(z-t)+g(z+t)+2\}^2-1\right]\,,\quad
(t>0,\, z>0)
\,,\end{equation} 
and $\phi(-z,t)=-\phi(z,t)$, where
\begin{equation} 
g(y)=\sqrt{f(h(y))}-1\,,\quad\mbox{for } y>0\,,
\end{equation} 
and $g(-y)=-g(+y)$ (See Fig.~4).

\begin{figure}[ht]
\begin{center}
\includegraphics[width=6cm]{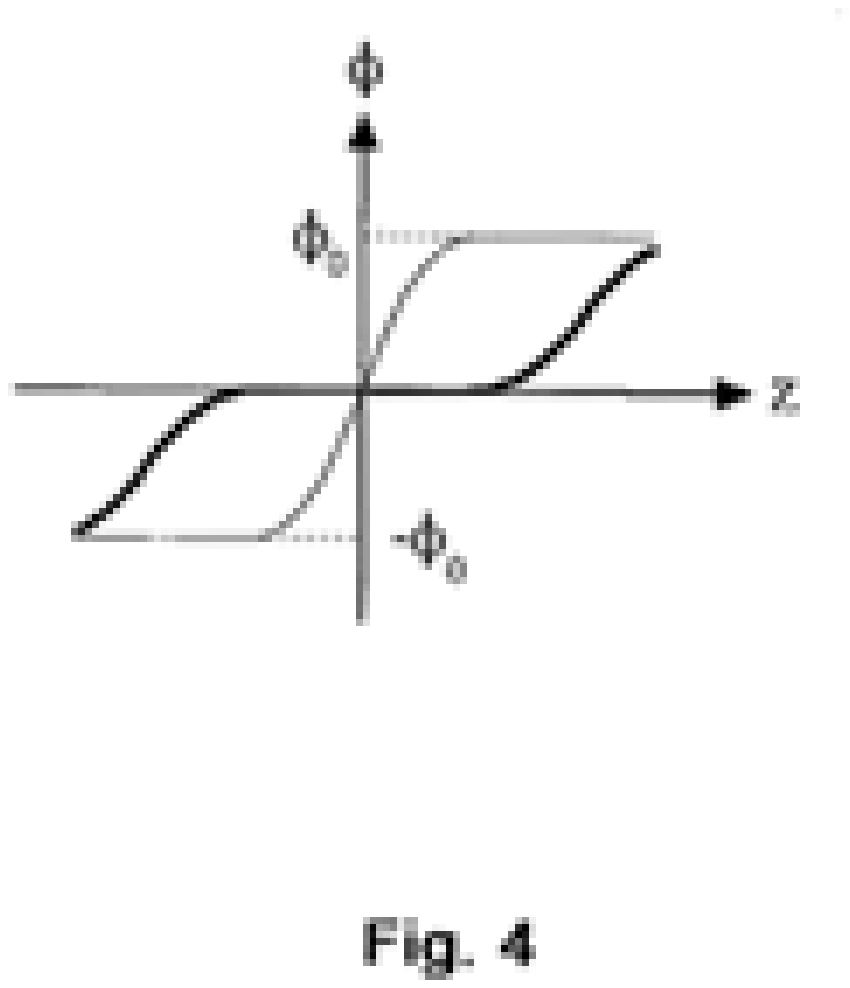}
\caption{The breakup of plane-symmetric walls in the true vacuum. The
thick line indicates the configuration og $\phi$ at $t=2z_0$.}
\label{f4}\end{center}
\end{figure}

We find that the walls separated from each other soon become to
move at nearly light speed.

Two essential characteristics of the behaviour of the wall after
$\lambda$ disappears are: (i) the energy stored at the wall cannot
dissipate suddenly; (ii) the bulk of energy cannot stay still, according
to the equation of motion.

Now, we consider the significance or the domain wall in our model in
cosmology.

The only first-order phase transition in the universe ever known is the
(confinement) transition of quantum chromodynamics (QCD) \cite{16}.
Although the order of the phase transition is yet debatable, we presume
the first-order transition. The temperature of the phase transition is
considered as $T=\Lambda_{QCD}\approx 200 \mbox{MeV}$. Consecquently,
the energy density of the false vacuum, that is, the effective
cosmological term $\lambda$ in our context, is the order of
$\Lambda_{QCD}^4$.

When the temperature of the universe decreases below
$T_C=\Lambda_{QCD}\approx 200 \mbox{MeV}$, the false vacuum energy
dominates over the thermal energy of radiations and particles. At the
same time the scalar field $\phi$ takes the form of the wall due to
non-linear self-coupling, since the thermal fluctuations of the field
become sufficiently small. The domain walls are initially generated by
the Kibble mechanism \cite{17}. We expect at least one wall in each
horizon-size volume.

We must check the consistency of calling the object as ``wall''. If the
thickness of the ``wall'' is wider than the horizon scale of the
universe at that time, we can no longer can it as a ``wall''. Moreover,
the process of formation of the object becomes closely dependent on the
growth of the fluctuation of the field. The ratio of the thickess of the
``wall'' to the hoizon length is
\begin{equation}
\frac{\mbox{thichness
of the
``wall''}}{\mbox{horizon
length}}=\frac{z_0}{R_H}=\frac{\sqrt{\frac{\phi_0^2}{\lambda(\delta
f)}}}{\sqrt{\frac{1}{\kappa^2\lambda}}}=\sqrt{\frac{\kappa^2\phi_0^2}{\delta
f}}\,.
\end{equation}
We find there is no contribution as long as the following inequalities
are satisfied
\begin{equation}
\kappa^2\phi_0^2\ll\delta f\ll\kappa\phi_0\ll 1\,,
\label{(9)}
\end{equation}
which we obtain by combining the cohd1tiort adopted when we led the
approxmate solution of the ``wall''. Note that the conditions are
independent of the value of $\lambda$. At the same time, We can safely
ignore the effect of the expansion of the universe.

The mass density of the wall $\eta$ is given by
\begin{equation}
\eta\simeq\sqrt{\lambda(\delta f)}\cdot\phi_0\simeq(\delta f)^{1/2}
\Lambda_{QCD}^2
\phi_0\,.\label{(10)}
\end{equation}
Along with progress of the phase transition, bubbles of the true vacuum
are born and expand \cite{18}. The phase transition is over when all
the spatial regions come to belong to the true vacuum as a result of the
percolation of the bubbles. In the true vacuum where the effective
cosmological constant is nearly zero, the domain wall made of the
configuration of $\phi$ decays. Therefore the wall begins to break
inside the bubbles. Consequently, the evolution of fragments of walls
depends on the manner of bubble nucleations. Even if the domain walls are
initially plane-symmetric, they take random shape at the end of the phase
transition because of randomness in nucleation of bubbles.

If the wall formed during QCD phase transition remains to be static
after the transition, the existence of wall conflicts with cosmological
observation \cite{19} as in the case of axionic domain walls \cite{20}.
The wall in our model, however, completely splits off after phase
transition and the remnants move away by almost light speed. Thus we can
avoid the above-mentioned problem; it can be said that the universe is
fiIled with the relativistic gas of walls \cite{21}. Let us estimate the
present mass density of the relics of the wall. At the present time, the
density $\rho$ is given as
\begin{equation} 
\rho_w\ge\frac{\eta
R_H^2}{R_H^3}\left(\frac{T_{present}}{T_c}\right)^\alpha\,,
\end{equation} 
where $T_{present}\approx 2.75 \mbox{K}\approx 0.24 \mbox{meV}$ and
$T_C\approx \Lambda_{QCD}$.

Here the parameter $\alpha$ $(3\le\alpha\le 4)$ is required for the
reasm that we take an account of the effect of transformation of the
energy of fragments into that of other massive particle. The constraint
that the energy density $\rho_w$ is lower than the critical density of
the present universe
$(\rho_c\sim 2.6\times 10^4 T_{present}^4)$ yields
\begin{equation} 
\sqrt{(\delta
f)\kappa^2\phi_0^2}\left(\frac{T_{present}}{\Lambda_{QCD}}\right)^{\alpha-4}\le
10^4\,.
\label{(12)}
\end{equation} 

This constraint can easily be satisfied for a wide range of parameters.

The contribution of $\rho_w$ to the total energy density might explain
the missing mass in the universe, which has not yet been identified. It
is dependent on a hidden sector coupled to $\phi$ whether the relics of
the wall can provide the dark matter which gathers other matters to
generate galaxies. This is because the fragments of wall remain to
move fast if there is no interaction between $\phi$ and other field. In a
class of extended inflationary scenario, hidden sectors which violate the
weak equivalence principle have been considered \cite{10,11}. In our
context, we have no compelling ground to determine the property of the
hidden sector. Since the stage of our investigation in this paper is
primitive, we will examine the property of other fields coupled to $\phi$
and gravity in a future work including numerical simulations of walls.

In the rest of the present paper, we investigate the connection to the
oscillating universe model \cite{22,23}.

In the extended model we treated, $f(\phi)$ was not a smooth function,
for a simple analysis. Of course, this is not an essential thing. From
now on, we suppose $f$ takes the form of the trigonometric function
(Fig.~1), since we want to consider a coherent oscillation of the field
$\phi$. Our analysis below need only a rough sketch of $f$; dimensional
analysis can lead to a sufficient estimation.

Suppose that there exists a non-vanishing (positive) cosmological
constant in the present universe. Then an effective interaction of the
scalar $\phi$ appears, provided that the value of the cosmological term
is of the same order of the thermal energy or more. In this time, we
consider a spatially homogeneous, oscillating field. Using an expansion
of the effective interaction with respect to small $\phi$, we obtain the
period of oscillation as
\begin{equation} 
\tau\approx\sqrt{\frac{\phi_0^2}{\lambda_p (\delta f)}}\,,
\label{(13)}
\end{equation} 
where $\lambda_p$ is the present cosmological constant.

Recent astronomical observation revealed very large, periodic
structure with a period of about $400$ million lightyears in the
universe \cite{24}. Morikwa and serveral authors have claimed that the
structure is an illusion induced from an error in estimation of the
distance scale by means of measurement of galaxies'
redshift \cite{22,23}. An assumption that the expansion rate of the
scale factor is oscillating leads to such an artificial pattern.
Morikawa took a model in which the oscillation of a coherent scalar field
bring about the oscillation of ``Hubble constant'' through modified
Einstein equations. Here we try to identify such a scalar field with
$\phi$ in our model.

The period (\ref{(13)}) becomes $400$ million years when we take
$\lambda_p$ as
\begin{equation} 
\lambda_p=\frac{\kappa^2\phi^2}{\delta f}\times 10^{-8} (\mbox{eV})^4\,.
\label{(14)}
\end{equation} 

According to another group of authors in Ref.~\cite{23}, a finite
cosmologlcal constant of order of $\rho_c\approx(3 \mbox{ meV})^4$ is
demanded by observations of distant galaxies. If we take such an allowed
value for $\lambda_p$ in (\ref{(14)}),
\begin{equation} 
\frac{\kappa^2\phi^2_0}{\delta f}=8\times 10^{-3}\,.
\label{(15)}
\end{equation} 
If we further require a sufficient amplitude of oscillating Hubble
parameter which can explain the observation, $\delta f$ must be larger
than $3\times 10^{-3}$, according to Steinhardt in Ref.~\cite{25}. If we
choose $\delta f=3\times 10^{-3}$, we must set $\phi_0=10^{15}
\mbox{GeV}$ according to (\ref{(15)}). These values for parameters give
rise to no contradiction with the previous constraints (\ref{(9)}). The
constraint (\ref{(12)}) is also cleared $\alpha>3.4$. This value of
$\alpha$ means that a finite fraction of energy density produced by the
domain walls remains in the form of relativistic gas.

It can never be justified that our naive model describes the true nature
of the gravity theory. The time variation of the gravitational constant
and other constant is strictly constrained by observation; even if the
constraints are able to be satisfied, fine-tuning is necessary, for
instance, in the phase of oscillations. The important point we want to
claim is that a general modification of gravity could explain different
problems in cosmology in the same time, such as dark matter problem and
problems in large scale structure of the universe. Accordingly, it is
interesting and significant to investigate the detail of the consequence
of the model and to make more complicated models which have more
relevance to cosmological puzzles and no conflicts with observation. We
also have to investigate the possibility that the first-order transition
should be identified to unknown transition, for instance, in relation
to sub-quark dynamics.

We wish to study the effect of cosmic expansion and gravity on the
production of walls as well as the numerical simulation including the
bubble nucleation and the wall intersection.

After completing this work, the present author became aware of the
paper \cite{26}. They also considered the wall which disappears after a
certain phase transition. Moreover they claimed that the wall can
provide the seed for generating galaxies. While their motivation in
cosmology is very interesting, their analysis depends crucially on a
peculiar model. We think that we must consider models which are related
with many cosmological aspects and can be checked by observations.

\section*{Acknowledgments}
 The author would like to thank T. Suzuki and T. Maki for some
useful comments. He would also like to thank A. Sugamoto for reading
this manuscript.

 KS is indebted to Soryuusi shogakukai for financial support. He also
would like to acknowledge financial aid of Iwanami F\=ujukai.


\end{document}